# The directional nature of hydrophobic interactions: Implications for the mechanism of molecular recognition

Qiang Sun[*]

*Key Laboratory of Orogenic Belts and Crustal Evolution, Ministry of Education, The School of Earth and Space Sciences, Peking University, Beijing, 100871, China*

**ABSTRACT:**

Based on recent studies on hydrophobic interactions, it is devoted to investigate the directional nature of hydrophobic interactions. It means that the hydrophobic interactions are dependent on the relative orientations as the solutes tend to be aggregated in water. In H1w process, they are attracted to approach each other in the specific direction with lower energy barrier until their surfaces become contact. In H2s process, to maximize the hydrogen bondings of water, the solutes are aggregated in the specific direction to minimize the ratio of surface area to volume of them. Additionally, with decreasing the separation between them, the short-range interactions between the solutes become stronger. In addition, these can be demonstrated by the calculated potential of mean force (PMF) using molecular dynamics simulation. From this work, the hydrophobic driven model is proposed to understand the specificity and affinity of molecular recognition in water.

[*] Corresponding author
E-mail: QiangSun@pku.edu.cn

# 1. Introduction

The hydrophobic effects, generally described as the tendency of non-polar molecules (or molecular surfaces) to be aggregated in an aqueous solution, are involved and believed to play an important role in many physical, chemical and biological processes, especially for molecular recognition. It refers to the process in which two (or more) molecular binding partners interact with each other through non-covalent interactions to form a specific complex. Currently, three basic models have been proposed to explain the molecular recognition, such as lock-and-key [1], induced fit [2] and conformational selection models [3-6]. To understand the mechanism of molecular recognition, it is necessary to investigate the hydrophobic interactions.

To unravel the physical origin of hydrophobic effects, many works have been carried out [7-16]. The classic mechanism of hydrophobic interactions, proposed by Frank and Evans [7], advanced by Kauzmann [8], Tanford [9] and others, was based on "iceberg" model. It means that, in comparison with bulk water, water around hydrophobic surface becomes more orderly-perhaps more "ice-like" or "clathrate-like" [7]. When two nonpolar surfaces are associated with each other, ordered molecules of water solvating each surface may be released to the bulk, which yields a favorable change in entropy (ordered water becomes less ordered). Therefore, the hydrophobic association is usually regarded to be driven by an increase in entropy. In contrast, for a sufficiently large solute, water dewets the solute surface, and the solute-water interface is similar to that between vapor and liquid water. Therefore, as suggested by Stillinger [10], the hydrophobic interaction of large-scale hydrophobic solutes may be different from small scale ones. According to Lum, Chandler, and Weeks (LCW) works [13-16], they provided a quantitative description of structural and thermodynamic aspects of hydrophobic hydration over the entire small to large length scale region. It can be expected that the small to large crossover length scale occurs at the nanometer length scale [15,16]. Additionally, the overall hydration free energy changes from growing linearly with the solvated volume to growing linearly with the solvated surface area [15,16].

As a solute is dissolved into water, the interface can be expected to occur between the solute and water, which undoubtedly affects the structure of water. Based on the structural studies on water and air/water interface [17-20], hydration free energy is derived, and utilized to investigate the origin of hydrophobic interactions [21-25]. In our recent studies [21,22], with reference to Rc (critical radius), it can be divided into initial and hydrophobic solvation processes. Additionally, with increasing solute concentrations, various dissolved behaviors of solutes can be expected in different solvation processes, such as dispersed and accumulated distributions in water. For the origin of hydrophobic effects, it is due to the competition between the hydrogen bondings of bulk water and those of interfacial water [21,22].

In hydrophobic solvation process, with reference to $R_H$ (hydrophobic radius), it can be divided into H1w and H2s hydrophobic processes [22]. In this work, based on our recent studies [21,22] on hydrophobic interactions, it is devoted to investigate the directional nature of hydrophobic interactions. This means that, as the solutes are attracted to be accumulated in water, hydrophobic interactions are dependent on the relative orientations between them. Additionally, various directional natures of hydrophobic interactions can be expected in H1w and H2s processes, respectively. In addition, potential of mean force (PMF) is calculated using molecular dynamics simulations to demonstrate the directional nature of hydrophobic interactions. From this, it can be extended to understand the mechanism of molecular recognition.

## 2. Molecular dynamics
### 2.1. Molecular dynamics simulations

The molecular dynamics simulations were carried out using the program NAMD 2.12 [26]. The simulations were conducted on different systems, each containing a target solute (graphene sheet) and a test solute ($C_{60}$ fullerene). To investigate the directional nature of hydrophobic interactions, the graphene sheet was fixed, and the test fullerene was constrained to approach it in different directions. Additionally, a few simulations were also conducted to investigate the kinetic pathway as the $C_{60}$ fullerene was associated with the graphene in water.

In this work, the CHARMM (Chemistry at Harvard Macromolecular Mechanics) force field [27] was utilized to describe the interatomic interactions. The water molecules were simulated using the intermolecular three point potential (TIP3P) water model. The electrostatic interactions were calculated using the particle mesh Ewald (PME) algorithm. Additionally, non-bonded van der Waals interactions were smoothly switched to zero between 10 and 12 Å.

The molecular dynamics simulations were carried out in the isobaric-isothermal ensemble (NPT). The simulated temperature was 300 K, employing moderately damped Langevin dynamics.

The pressure was maintained at 1 atm using a Langevin piston. The initially simulated box was 40 Å×40 Å×90 Å. In addition, periodic boundary conditions were utilized in the three directions of Cartesian spaces. For each simulation, the simulated time was 6 ns, and the equations of motion were integrated with a time step of 2 fs.

**2.2. ABF calculations**

For the directional nature of hydrophobicity, it means that the strength of hydrophobic interactions may be dependent on the relative orientation between the solutes as they are aggregated in water. To investigate the directionality of hydrophobic interactions, potential of mean force (PMF) is calculated using NAMD with the adaptive biasing force (ABF) [28-33] extensions integrated in the Collective Variables (CVs) module [34].

The ABF method [28-33] is a combination of probability density and constraint force methods. It is based on the thermodynamic integration of average force acting on generalized reaction coordinate ξ. In ABF calculations, to oppose the actual force arising from simulated system, a biasing force is periodically applied to the coordinate in order to generate what is effectively a random walk (purely diffusive dynamics).

In ABF calculations, the reaction coordinate is usually divided into several discrete bins, and the average force can be determined as [31-33],

$$F_\xi\left(N_{step},k\right) = \frac{1}{N\left(N_{step},k\right)} \sum_{i=1}^{N\left(N_{step},k\right)} F_i(t_i^k) \quad (1)$$

where $N(N_{step},k)$ is the number of samples collected in the bin k after $N_{step}$ simulation steps, $F_i(t_i^k)$ is the computed force at iteration i, and $t_i^k$ is the time at which the i*th* sample was collected in the bin k. Then, the free-energy difference, $\Delta A_\xi$, between the end point states can be determined through summing the force estimates in individual bins,

$$\Delta A_\xi = -\sum_{i=1}^{M} \overline{F_\xi}\left(N_{step},k\right) \delta\xi \quad (2)$$

In this study, the target solute (graphene sheet) is fixed, and the test solute ($C_{60}$) is restrained to move to the target solute in different direction along Z axis. In the ABF calculations, the distances between the solutes are regarded as CVs, and are broken down into several consecutive windows, each with 2.8-3.0 Å wide. From these, PMFs are determined, and can be utilized to investigate the directional nature of hydrophobic interactions.

## 3. Discussions

### 3.1. Hydrophobic interactions

From Einstein's words [35], "A theory is the more impressive the greater the simplicity of its premises is, the more different kinds of things it relates, and the more extended is its area of applicability. Therefore the deep impression which classical thermodynamics made upon me. It is the only physical theory of universal content concerning which I am convinced that within the framework of the applicability of its basic concepts, it will never be overthrown."

In thermodynamics, Gibbs free energy, $\Delta G = \Delta H - T \cdot \Delta S$, is a thermodynamic potential that measures the capacity of a thermodynamic system to do maximum (or reversible) work at a constant temperature and pressure. It is one of the most important thermodynamic quantities to characterize the driving forces. Thermodynamically, the enthalpic component ($\Delta H$) quantifies the total energy of a thermodynamic system, and the entropic component ($\Delta S$) quantifies the change in disorder of the overall system. For any spontaneous process, the change in Gibbs free energy of the system is negative when the system reaches an equilibrium state at constant pressure and temperature. Therefore, all chemical systems tend naturally toward states of minimum Gibbs free energy.

In fact, as the solutes are embedded into water, the total Gibbs free energy can reasonably be expressed as follows,

$$\Delta G = \Delta G_{Solute-solute} + \Delta G_{Solute-water} + \Delta G_{Water-water} \qquad (3)$$

where $\Delta G_{Solute-solute}$ is related to the interactions between the solutes, $\Delta G_{Solute-water}$ is due to the interactions between the solutes and water, and $\Delta G_{Water-water}$ is the Gibbs energy of bulk water, respectively. Regarding to the van der Waals interactions between solutes, they are short-range forces, and only interactions between the nearest particles need to be considered. To understand the driving force to make the solutes to be accumulated in water, it is necessary to investigate the structure of water, and the effects of dissolved solutes on water structure.

To investigate the structure of liquid water, many experimental and theoretical works have been carried out. Various structural models of water have been proposed, they can roughly be partitioned into two categories, (a) mixture models, and (b) continuum models (or distorted hydrogen bonding) [36]. The OH vibrations are sensitive to hydrogen bondings, and widely applied to investigate the water structure. According to Raman spectroscopic studies [17-19], when three-dimensional hydrogen bonding occurs, the OH vibrations are mainly dependent on local hydrogen bondings (the first shell) of a water molecule, and the effects of hydrogen bondings beyond the first shell on OH vibrations are weak. Therefore, various OH vibrational frequencies can be ascribed to OH vibrations engaged into different local hydrogen bondings.

At ambient conditions, the Raman OH stretching bands of water can reasonably be deconvoluted into five sub-bands, and each sub-band is ascribed to OH vibrations engaged into various local hydrogen bondings, such as DDAA (double donor-double acceptor, tetrahedral), DDA (double donor-single acceptor), DAA (single donor-double acceptor), DA (single donor-single acceptor), and free OH vibrations, respectively [17-19]. For ambient water, it can be derived that a water molecule interacts with the neighboring (the first shell) water molecules through various local hydrogen bondings [19]. Additionally, the structure of water may be influenced by temperature, pressure, dissolved salts, and confined environments, which will be rearranged to oppose the changes of external conditions.

As the solute is dissolved into water, the interface can be expected to appear between the solute and water, which undoubtedly affect the structure of liquid water. The OH vibration is mainly dependent on the local hydrogen bondings of a water molecule, therefore the solute mainly affect the structure of interfacial water (the topmost water layer at the solute/water interface) [20]. In fact, this can be demonstrated by recent experimental studies [37-42] on the structure and dynamics of water around the dissolved salts, which means that the effects are largely limited to the first solvation shell. In comparison with bulk water, no DDAA (tetrahedral) hydrogen bondings can be expected in the interfacial water [20]. Therefore, the Gibbs free energy of interfacial water, incurred by the dissolved solute, can be expressed as,

$$\Delta G_{Solute-water} = n_{DDAA} \cdot \Delta G_{DDAA} \cdot R_{Interfacial\ water} \qquad (4)$$

where $n_{DDAA}$ is the hydrogen bondings per water molecule of DDAA hydrogen bondings, $\Delta G_{DDAA}$ is the Gibbs energy of DDAA hydrogen bondings, and $R_{Interfacial\ water}$ is the ratio of interfacial water to bulk water.

In thermodynamics, the process of embedding a hydrophobic solute into water is equivalent to that of forming an empty spherical cavity in water. After the solute is treated as a sphere, the molecular number ratio of the interfacial water layer to volume can be determined to be $4 \cdot r_{H2O}/R$, where $r_{H2O}$ is the average radius of a $H_2O$ molecule, and R is the radius of the solute. From this, the hydration free energy can be determined as,

$$\Delta G_{Hydration} = \Delta G_{Water-water} + \Delta G_{Solute-water} = \Delta G_{Water-water} + \frac{8 \cdot \Delta G_{DDAA} \cdot r_{H2O}}{R} \qquad (5)$$

The hydration free energy is the sum of Gibbs energy of interfacial water and bulk water. Thermodynamically, the lower Gibbs free energy, the more stable system. Therefore, as a solute is dissolved into water, the structural transition can be expected as,

$$\Delta G_{Water-water} = G_{Solute-water} \qquad (R = Rc) \qquad (6)$$

where Rc is the critical radius of solute. At 293 K and 0.1 MPa, the Rc is determined to be 6.5 Å. Based on our recent works [20], with reference to Rc, it can reasonably be divided into the initial and hydrophobic solvation processes, respectively (Figure 1).

The Gibbs energy of interfacial water ($\Delta G_{Solute-water}$) is inversely proportional to the radius of the solute, or proportional to the ratio of surface area to volume of solute. Therefore, with increasing solute concentrations, various dissolved behaviors of solutes can be expected in different solvation processes. In the initial solvation process ($\Delta G_{Solute-water} < \Delta G_{Water-water}$), it is dominated by the Gibbs energy of interfacial water, and the solutes tend to dispersed in water. It seems that there exist "repulsive" forces between the solutes as they are pushed together. However, in hydrophobic solvation process ($\Delta G_{Solute-water} > \Delta G_{Water-water}$), the solutes are expected to be accumulated in water to maximize the bulk water. It seems that there exist the "attractive" forces between them. Additionally, these can be demonstrated by our recent studies on the dependence of hydrophobic interactions on the solute size [22].

Regarding to the origin of hydrophobic interactions, it is due to the structural competition between the hydrogen bondings of interfacial water and those of bulk water. Therefore, hydrophobic interactions can reasonably be regarded as the "effects" rather than "bond". Generally, hydrophobic effects mean the tendency of the non-polar molecules (or molecular surfaces) to be accumulated in solutions. From our works [21], the hydrophobicity is reasonably described as the tendency of the minimization of the ratio of surface area to volumes of solutes to maximize the hydrogen bondings of water. In fact, hydrophobic interactions may be due to the hydrogen bonding of bulk water is stronger than that of interfacial water. Therefore, hydrophobic effects may be extended to other systems only if the strength of solute/solvent is weaker than bulk solvent.

As the solutes are dissolved into water, they mainly affect the structure of interfacial water. Due to hydrophobic interactions, the solutes are attracted to approach each other in order to maximize the hydrogen bondings of water. It can be derived that, as no water molecules can be found between them, the surfaces of solutes become contact, which undoubtedly decreases the solute surface to be available for interfacial water. In our recent works [22], as the solute surfaces become contact, the corresponding separation between them can be termed as $R_H$ (hydrophobic radius). During the accumulation of dissolved solutes in water, the Gibbs free energy of interfacial water can reasonably be expressed as,

$$\Delta G_{Interfacial\ water} = \gamma \cdot \Delta G_{Solute-water} \qquad (7)$$

where $\gamma$ is termed as the geometric factor. It is utilized to reflect the changes of solute surface during the aggregation of solutes in water. In fact, as the solutes are accumulated in water, it may be also accompanied with the changes of solute volume. Therefore, $\gamma$ is generally defined as,

$$\gamma = \frac{\left(Surface\ area/Volume\right)_{Aggregate}}{\left(Surface\ area/Volume\right)_{Non-aggregate}} = f\left(1/r_{Separation}\right) \qquad (8)$$

where $r_{Separation}$ is the separation between the solutes. In our recent works [22], with reference to $R_H$ (or $\gamma$), it is divided into H1w and H2s hydrophobic solvation processes, respectively.

In hydrophobic solvation process, with decreasing the separation between the solutes, the water molecules between the solutes are undoubted expelled into bulk water until the surfaces of solutes are accumulated in water. Therefore, various hydrophobic interactions can be expected in H1w and H2s hydrophobic processes, respectively. In addition, as the separation between the solutes being less than $R_H$, the short-range solute-solute interactions, such as van der Waals force, etc. become important. Therefore, the dissolved solutes are affected by not only the hydrophobic interactions but also the short-range forces.

In H1w hydrophobic process (>$R_H$), the water molecules are expected to exist between the solutes, therefore $\gamma$ is equal to 1. To maximize the hydrogen bondings of water, the solutes are attracted to approach each other in water. With decreasing the separation between the solutes, the water molecules between them are expelled into bulk water. Therefore, the energy barriers can be expected in H1w hydrophobic process, which is related to the expelled water molecules. To be more thermodynamically stable, the solutes are expected to approach in the specific direction with lower energy barrier, in which less water molecules between the solutes are expelled into bulk water. Therefore, hydrophobic interactions are directional in H1w process.

In H2s hydrophobic process (<$R_H$), the surfaces of dissolved solutes begin contact. Due to the aggregation of solute surface, this decreases the surface of solutes to be available for interfacial water. From the above, the Gibbs free of interfacial water is proportional to the ratio of surface area to volume of solutes. To be more thermodynamically stable, this leads the solutes to be accumulated in the specific direction to minimize the ratio of surface area to volume. In fact, this can be demonstrated by molecular dynamics simulations on Graphite sheets [43,44], and CNT [44,45]. Therefore, the directional nature of hydrophobic interactions can also be found in H2s process. Of course, it is different from the directionality in H1w hydrophobic process.

When the solutes are embedded into water, they mainly affect the structure of interfacial water. Due to the long-range hydrophobic interactions, the solutes can be expected to be accumulated in hydrophobic solvation process. Because of the directional nature of hydrophobic interaction, it can be derived that the solutes are attracted to approach each other in the specific direction to be more thermodynamically stable. In H1w hydrophobic process, the solutes approach in the direction with lower energy barrier, which is related to the expelled water from the intermediate between the solutes into bulk water. However, during H2s hydrophobic process, the solutes become contact, and are accumulated to minimize the ratio of surface area to volume. In addition, it should be noted that the short-range intermolecular interactions, such as van der Waals interactions, etc, become stronger in H2s process. From the above, this may be applied to investigate the mechanism of molecular recognition.

**3.2. Molecular dynamics simulations**

Due to the directional nature of hydrophobic interactions, the hydrophobic interactions are dependent on the relative orientation between the solutes as they are attracted to approach each other in solutions. In other words, this indicates that, as the test solute is restrained to move to the target solute in fixed direction, the hydrophobic interactions are related to the geometric shape of them. In this study, to investigate the directional nature of hydrophobicity, the graphene sheet is fixed, the $C_{60}$ fullerene is constrained to move to the different site of sheet, such as the center, the edge, and the corner of the plane (Figure 2). Therefore, as the $C_{60}$ fullerene approaching the sheet in different direction, various ratio of surface area to volume of them can be obtained. After the PMFs are calculated, these can be utilized to investigate the directional nature of hydrophobic interaction.

As the fullerene approaches the graphene sheet, the PMFs can be determined through ABF calculations (Figure 3). Three minima can be found in the calculated PMFs. These are in accordance with other calculated PMFs during the aggregation of two fullerenes, two graphene planes, and two CNTs in water [22,43-45]. The first minimum is the contact minimum. The second minimum is attributed to the solvent-separated PMF, which means that only one water molecular layer can be found between the solutes. In addition, a third minimum is also observed, which corresponds to a double water molecular layer located between them (Figure 3).

As the solutes are attracted to approach each other, obvious energy barrier can be found as a double water molecular layer is located between them (Figure 3). In fact, this is in correspondence with two interfacial water layers to be located between the solutes. Therefore, this indicates that the dissolved solutes mainly affect the structure of interfacial water, which is in agreement with the above discussion on the effects of solute on water structure [18]. From the calculated PMFs, with decreasing the separation between the solutes, the energy barriers can be found between neighboring minima, which are associated to the expulsion of a single water layer in the confined volume as the solutes move closer to each other.

In addition, although the three minima can be observed in the calculated PMFs, obvious difference can be found especially at the height of energy barriers (Figure 3). When $C_{60}$ approaches the plane in different directions, the energy barriers are dependent on the relative orientation between them, which can be listed as, (Energy barrier)$_{C60-center\ of\ plane}$>(Energy barrier)$_{C60-edge\ of\ plane}$>(Energy barrier)$_{C60-angle\ of\ plane}$. In fact, with decreasing the separation between the solutes, water molecules between them are undoubtedly expelled into bulk water. The number of expelled water molecules may be dependent on the relative orientations between the solutes. Therefore, this indicates that the energy barrier may be closely related to the expelled water during the association of solutes in water.

To investigate the water-induced contributions in the process of solute association, the PMFs between solutes under a vacuum are also calculated (Figure 3). Hence, the water-induced contributions to the PMFs can be determined as,

$$\Delta G_{Wter-induced} = \Delta G_{Solute-solute\,in\,water} - \Delta G_{Solute-solute\,in\,vacuum} \qquad (9)$$

Therefore, they can be applied to investigate the directional nature of hydrophobic interactions as the fullerene approaching the graphene plane in different directions.

In fact, with decreasing the separation between the solutes, the solute surfaces are expected to become contact as no water molecules can be found between them ($<R_H$). Due to the aggregation of solute surface, this decreases the surface area of solutes to be available for interfacial water. From the simulations, the $R_H$ can be determined to be 7.64 Å, 18.53 Å, and 22.71 Å as $C_{60}$ moving to the center, the edge, and the corner of the Graphene sheets (Figure 3). With reference to $R_H$ (or $\gamma$), it can be divided into H1w and H2s hydrophobic processes, respectively. From the above discussion on hydrophobic interactions, various hydrophobic interactions can be expected in H1w and H2s processes. Regarding to the fullerene perpendicularly moving to the center of graphite plane, the calculated $\Delta G_{Water-induced}$ can reasonably be fitted as the following (Figure 4),

$$\Delta G_{Water-induced} = a + \gamma \cdot b/(r - 6.6) \qquad (10)$$

where r-6.6 is the separation between the $C_{60}$ and graphite, a and b are fitted to be -8.12 and 6.25, respectively.

As the $C_{60}$ fullerene approaching the graphene sheet in H1w process ($>R_H$), it can be found that the $\Delta G_{Water-induced}$ are dependent on the relative orientation between the solutes. From the calculated water-induced contributions (Figure 3), they are listed as, $\Delta G_{C60\text{-center of plane}} > \Delta G_{C60\text{-edge of plane}} > \Delta G_{C60\text{-angle of plane}}$ in H1w process. It can be found that the lower energy barrier can be expected as the fullerene moving to the corner of sheet. Therefore, the hydrophobic interactions are directional in H1w process. Additionally, this is in agreement with the above discussion on the height of energy barrier of PMFs (Figure 3). With decreasing separation between the solutes in H1w process, the water molecules between the solutes may be expelled into bulk. Regarding to the directional nature in H1w hydrophobic process, it may be closely related to the expelled water.

When the separation between the $C_{60}$ and the sheet being less than $R_H$, the surfaces of solutes begin to be accumulated, which decrease the surface area to be available for interfacial water. In H2s hydrophobic process, the dissolved solutes are accumulated to minimize the ratio of surface area to volume in order to maximize the hydrogen bonding of water. From the calculated water-induce contributions in H2s process (Figure 3), the hydrophobic interactions are listed as, $\Delta G_{C60\text{-center of plane}} > \Delta G_{C60\text{-edge of plane}} > \Delta G_{C60\text{-angle of plane}}$. Therefore, the hydrophobic interactions may be directional in H2s process. Additionally, the hydrophobic interactions can be expected to be strongest as the $C_{60}$ is associated with the center of graphene sheet.

The dissolved solutes mainly affect the structure of interfacial water. Therefore, as the solutes are embedded into water, it can be divided into the interfacial and bulk water. During the accumulation of solutes in hydrophobic solvation process, it leads to the transformation from interfacial water into bulk water. For the directional nature of hydrophobic interactions, it may be related to the transformation between them. In H1w hydrophobic process, as the $C_{60}$ fullerene moving to the corner of Graphene sheet, the lower energy barrier can be found (Figure 3), which is in accordance with the weaker transformation from interfacial to bulk water (Figure 5). However, in H2s hydrophobic process, the stronger hydrophobic interactions can be found (Figure 3) as the $C_{60}$ are associated with the center of the sheet, which is accompanied with the obvious transformation from interfacial to bulk water (Figure 5).

To investigate the origin of directional nature of hydrophobic interactions, the hydrogen bondings of water are calculated during the aggregation of the solutes in water. In this study, the geometrical definition of hydrogen bonding is utilized to determine the hydrogen bondings in water [46], which is determined by VMD [47] (Visual Molecular Dynamics) program. According to the geometrical definition, the hydrogen bonding is considered to exist between two water molecules if the oxygen-oxygen distance ($r_{OO}$) and ∠OOH angle between two water molecules are less than 3.5 Å and 30°, respectively.

From molecular dynamic simulations, the hydrogen bonding number (average number of hydrogen bonds per water molecule, nHB) can be calculated (Figure 6). In comparison with bulk water, due to the truncations of hydrogen bonding at the solute/water interface, the hydrogen bondings of interfacial water is less than bulk water (Figure 6). Therefore, hydrophobic interactions can be ascribed to the structural competition between the interfacial and bulk water. In hydrophobic solvation process, to maximize the hydrogen bondings, this leads to the solutes to be aggregated in water. Regarding to the driving force of solute association, it is reasonably ascribed to maximize the hydrogen bondings of water.

As the separation between the solutes being less than $R_H$ in H2s process, no water molecules can be found between them, and the solute surfaces become contact in aqueous solutions. From the calculated PMFs, it can be found that the short-range solute-solute interactions, such as van der Waals force, are stronger than the hydrophobic interactions. Therefore, different form H1w process, the association between the solutes in H2s process is affected by not only hydrophobic interactions but also the solute-solute interactions. Additionally, to improve the stability of solute association, this can be achieved by increasing the strength of the solute-solute interactions.

In addition, similar to liquid-gas phase transition, dewetting transition process can be observed during the accumulation of solutes in water (Figure 5). In fact, water dewetting from non-polar confinement has been observed in a number of simulation studies relating to nanotubes and plates [48,49], water-protein interfaces [50], and collapsing polymers [51]. In combination with our recent studies [22], it can be derived that the dewetting is accompanied with the single-layer water molecules between the solutes are expelled into bulk water. Therefore, it may be closely related to the transition from H1w to H2s hydrophobic process.

When the solutes are embedded into water, due to the long-range hydrophobic interactions, they are attracted to approach each other until they are affected by the short-range molecular force. From the simulations, as the $C_{60}$ approaching the graphene sheet, the hydrophobic interactions are dependent on the relative orientation between them. In H1w process, the lower energy barrier can be expected as the $C_{60}$ moving to the corner of sheet. In H2s process, the stronger hydrophobic interactions can be found as the fullerene is associated with the center of plane. Additionally, the association of solutes is also affected by the short-range force in H2s process. From these, it may be applied to investigate the kinetic pathway during the $C_{60}$ fullerene is associated with the graphene sheet.

In this work, to investigate the kinetic pathway during the accumulation between $C_{60}$ and graphene sheet, the unbiased simulations are also conducted using NAMD program [26]. When the $C_{60}$ fullerene approaches the graphite plane, the lowest energy barrier can be found as the fullerene approaching the angle of graphite plane in H1w process (Figure 3). Therefore, it can be derived that this may be the pathway where the $C_{60}$ must go through in order to be aggregated with the graphene plane in water. As they are dissolved into water, due to hydrophobic interactions, they are attracted to move randomly and continuously adjust the motion direction. However, it should be noted that the $C_{60}$ must pass through the angle of plane before the solute surface of $C_{60}$ begins to be accumulated with the graphite plane (Figure 7). Additionally, as the surfaces of solutes are aggregated in H2s process, they tend to be accumulated to minimize the ratio of surface area to volume of them. Therefore, the $C_{60}$ may be associated with the center of the sheet to be more thermodynamically stable. Additionally, this can also be demonstrated by the molecular dynamics simulation using GROMACS [52,53] with OPLSAA force field and TIP4P water model (Supplementary).

According to the simulations on the hydrophobic interactions between $C_{60}$ and graphite in water, it is found that hydrophobic interactions may be directional, and which may be utilized to investigate the kinetic pathway during the association of solutes in water. In hydrophobic solvation process, it is divided into H1w and H2s process, respectively. In H1w process, the solutes are attracted to approach in the direction with the lowest energy barriers until the solute surface begins to be accumulated. In H2s hydrophobic process, due to the directional nature of hydrophobic interactions, the solutes can be expected to be accumulated in specific direction to minimize the ratio of surface area to volume of solutes. Additionally, with decreasing the separation between the solutes, the short-range force between the solutes, such as van der Waals interactions, etc. becomes stronger, which undoubtedly affects the stability of solute association (Figure 8).

## 4. Implications for molecular recognitions

Molecular recognition refers to the process in which two or more molecular binding partners interact with each other through non covalent interactions to form a specific complex. In especial, molecular recognition is fundamental to biology in that it governs signaling within and between cells, with prominent examples provided by the immune system, hormonal control of distant organs in higher organisms, and specificity of enzyme reactions [54]. Additionally, based on the understanding the mechanism of molecular recognition, it has direct applications in drug discovery and design.

To unravel the mechanism of molecular recognition, many experimental and theoretical works have been carried out. Three different models have been proposed, such as the Fischer's

"lock-and-key" [1], Koshland's "induced fit" [2], and "conformational selection" [3-6]. In the "lock-and-key" model, the conformations of the free and ligand-bound protein are essentially the same, while "induced fit" posits that conformational differences between these two states are the result of the binding interaction driving the protein towards a new conformation which is more complementary to its binding partner. According to recent experimental measurements, the "conformational selection" model is proposed, which postulates that all protein conformations pre-exist, and the ligand selects the most favored conformation. From this, a population shift is expected after binding the ensemble undergoes, which leads to the redistributing the conformational states. Therefore, regarding to the mechanism of molecular recognition, it still remains elusive.

Molecular recognition plays an important role in all biological processes. In fact, both specificity and affinity are required in the process of molecular recognition. The molecular specificity can be defined why two or more binding partners are allowed to approach and bind together [55-58], which is fundamental in various fields of molecular and biological sciences. For the binding affinity, it means the strength of these interactions, which determines whether an interaction will be formed in solution or not. Therefore, the binding affinity is closely related to the thermodynamic stability of the molecular association.

In fact, as the solutes are dissolved into water, various interactions can be expected to exist, such as those between water molecules, between solute and water, and between solutes, respectively. The solute-solute interactions, such as van der Waals force, etc. are the short-range interactions, which are closely related to the molecular association. Therefore, it is necessary that the solutes are firstly attracted to approach each other before they are affected by the short-range interactions between them. From these, it can be derived that molecular recognition, especially its specificity, may be closely related to the hydrogen bondings of water.

Based on our recent works [21,22], hydrophobic interactions are due to the competition between the hydrogen bonding of interfacial water and those of bulk water. Different from the interactions between solutes, hydrophobic interactions can be regarded to be a long-distance "attractive" force. From Hammer et al. experimental study [59], the effective range of the hydrophobic attraction can be found up to several micrometers. Therefore, the dissolved solutes may be attracted by the long-range hydrophobic interaction to approach each other until they are affected by the short-range force, such as van der Waals interactions, etc. For the molecular recognition, it can be described to be driven by the hydrophobic interactions.

From this work, it can be found that hydrophobic interactions may be directional. In other words, as the solutes are attracted to be accumulated in water, the strength of hydrophobic interactions is dependent on the relative orientations between the solutes. Additionally, various directional natures of hydrophobic interactions can be expected in H1w and H2s hydrophobic processes, respectively. In H1w process, with decreasing the separation between the solutes, the water molecules between them are expelled into bulk water. Therefore, the solutes tend to approach in the direction with lower energy barrier in H1w process. As the solute surfaces become contact in H2s process, the solutes are expected in the specific direction to minimize the ratio of surface area to volume of them. Due to the directional natures of hydrophobic interactions, these may be applied to understand the specificity of molecular recognition.

Regarding to the specificity of molecular recognition, it mean that a molecule can distinguish the highly specific binding partner from less specific partners to form a specific complex. The shape complementarity is regarded to play an important role in the association between protein and ligand, which states that, in order for a ligand to bind with reasonable affinity to a given target protein, it must fit in the cavity to which it binds without steric conflicts. From the above, this can be explained by the minimization of the ratio of surface area to volume of solutes in H2s hydrophobic process. In addition, to improve the specificity of molecular recognition, this can be achieved through decreasing the energy barrier in H1w hydrophobic process. This may be fulfilled by modulating the geometric characteristics of solute.

In fact, besides the specificity of molecular recognition, it can be found that hydrophobic interactions are also closely related to the stability of molecular association (affinity of molecular recognition). With increasing the strength of hydrophobic interactions, this also leads to the increase of the stability of solute association. In combination with our recent works [21] on hydrophobic interactions, this may be achieved by changing the geometric shape of solute, the size of solute.

From the simulations, the solute-solute interactions become stronger than hydrophobic interactions in H2s process. Regarding to the affinity of molecular recognition, it is related to not only the hydrophobic interactions but also the short-range interactions between the solutes, such as van der Waals force, etc. Therefore, this provides an approach to improve the stability of molecular association in the process of molecular recognition. In other words, the affinity of molecular recognition may be enhanced by the increase of the interactions between solutes, such as the Coulomb force, hydrogen bondings, etc.

In process of molecular recognition, the binding is intertwined processes involved in intra and inter-molecular interactions. For the macromolecules, they are inherently dynamics. Therefore, it is generally accepted that macromolecular interactions are rarely rigid and involve conformational changes. During the association of solutes in solutions, the conformational changes may be expected in order to be more thermodynamically stable. In fact, with increasing the molecular flexibility, it is helpful to improve both the specificity and affinity of molecular recognition.

In combination with our recent works [21], hydrophobic interactions may be closely related to the molecular specificity and affinity, which plays a central role in the process of molecular recognition. Therefore, it can be derived that molecular recognition may reasonably be regarded to be driven by the hydrophobic interactions (hydrophobic driven model) (Figure 8).

(I) As the solutes are dissolved into water, due to hydrophobic interactions, they are attracted by each other to approach until the solute surface begins to be aggregated. In H1w hydrophobic process, the solutes can be expected to approach in the specific direction with the lower energy barrier, which corresponds to the less expelled water molecules during the accumulation of solutes in water (Figure 8).

(II) In the H2s hydrophobic process, to maximize the hydrogen bondings of water, the solutes tend to be accumulated in the specific direction to minimize the ratio of surface area to volume of solutes (Figure 8). Therefore, various directional natures can be expected in H1w and H2s hydrophobic processes, respectively. This can be utilized to understand the specificity in the process of molecular recognition.

(III) With decreasing the separation between solutes in H2s process, the short-range interactions between solutes, such as van der Waals force, etc. become stronger (Figure 8). Regarding to the affinity of molecular recognition, it is related to both the long-range hydrophobic interactions and the short-range solute-solute interactions.

Hydrophobic interactions play an important role in the process of molecular recognition. Regarding to physical origin of hydrophobic interactions [21], it can reasonably be ascribed to the competition between the hydrogen bondings of bulk water and those of interfacial water. For the molecular recognition, it is closely related to the hydrogen bondings in water. It is well known that the hydrogen bondings in water are influenced by many factors, such as temperature, pressure, the addition of NaCl, and confined enviroments. Therefore, these undoubtedly affect the association of solutes in water.

**5. Conclusions**

Due to hydrophobic interactions, the dissolved solutes are expected to be accumulated in water. From this work, it can be found that hydrophobic interactions may be directional. This means that hydrophobic interactions are dependent on the relative orientations between the solutes. Additionally, it can be applied to understand the mechanism of molecular recognition.

(1) With reference to $R_H$, it can be divided into H1w and H2s hydrophobic processes. In H1w process ($>R_H$), the solutes are expected to approach in the direction with lower energy barrier. As the solute become contact in H2s process ($<R_H$), the solutes tend to accumulated in the specific direction to minimize the ratio of surface area to volume of them. Therefore, various directional natures can be expected in H1w and H2s processes.

(2) As the solutes are aggregated in water, with decreasing the solute-solute separation in H2s process, the short-range interactions between them become stronger. Regarding to the stability of association, it is related to not only the long-range hydrophobic interactions but also the short-range interactions between them.

(3) When the solutes are embedded into water, due to the hydrophobic interactions, they are expected to approach in the specific directions. As the surfaces of them become contact, they are affected by the short-range interactions, such as van der Waals force, etc. This can be applied to investigate the mechanism of molecular recognition (hydrophobic driven model), and understand the specificity and affinity during the molecular association.

**Acknowledgements**

This work is supported by the National Natural Science Foundation of China (Grant Nos. 41773050).

**Supplementary**

To investigate the kinetic pathway during the $C_{60}$ fullerene is associated with graphene sheet in water, MD simulations are carried out. This can be applied to understand the directional natures of hydrophobic interactions. The trajectories are shown in supplementary.

**Fig. 1.** Hydration free energy at 293 K and 0.1 MPa. In reference with Rc (the critical radius), it is divided into the initial and hydrophobic solvation processes, respectively.

**Fig. 2.** Different systems used to investigate the directional natures of hydrophobic interactions. Both initial (a-c) and final (d-f) configurations are shown.

**Fig. 3.** (a-c) The PMFs in water (red) and vacuum (blue) as $C_{60}$ fullerene approaching the graphite sheet in different directions, as shown in Fig. 2. The corresponding water-induced PMFs between the solutes are shown in (d-f).

**Fig. 4.** The water-induced PMF as the fullerene perpendicularly approaching the target plane at 300 K and 1 bar. In reference to $R_H$ (hydrophobic radius), it is divided into H1w and H2s hydrophobic processes. The strength of the hydrophobic interactions can be expressed as, $\Delta G = -8.12 + \gamma \cdot 6.25/(r-6.6)$. During the H1w process, $\gamma=1$. In the H2s process, the solute surface begins to be accumulated in water. The PMFs at various $\gamma$ (0.9, 0.7, 0.5) are drawn in squares.

**Fig. 5.** The changes of interfacial (a-c) and bulk (d-f) water as the fullerene approaching the plane in different directions, as shown in Fig.2.

**Fig. 6.** (a-c) The hydrogen bondings per water molecule of bulk (red) and interfacial (blue) water as the fullerene approaching the plane in different directions, as shown in Fig.2. The total hydrogen bondings are shown in (d-f).

**Fig. 7.** The kinetic pathway as the $C_{60}$ binding with the graphite plane. As the fullerene is associated with the graphite, it tends to pass through the angle of the plane (a-f). This is in accordance with the lowest energy barrier as the $C_{60}$ approaching the graphite, as shown in Fig. 2. Only interfacial water layers of the solutes are shown.

**Fig. 8.** The hydrophobic driven model of molecular recognition. The dissolved solutes mainly affect the structure of interfacial water (dashed line). Due to the long-range hydrophobic

interactions, the solutes are attracted to approach in the direction with lowest energy barrier in H1w process. As the solute surfaces begin to be aggregated in H2s process, to maximize the hydrogen bondings of water, they are accumulated in the specific direction to minimize the ratio of surface area to volume. Additionally, with decreasing the separation between solutes in H2s process, the short-range interactions between them become stronger. For the affinity of molecular recognition, it is related to both the hydrophobic interactions and the solute-solute interactions.

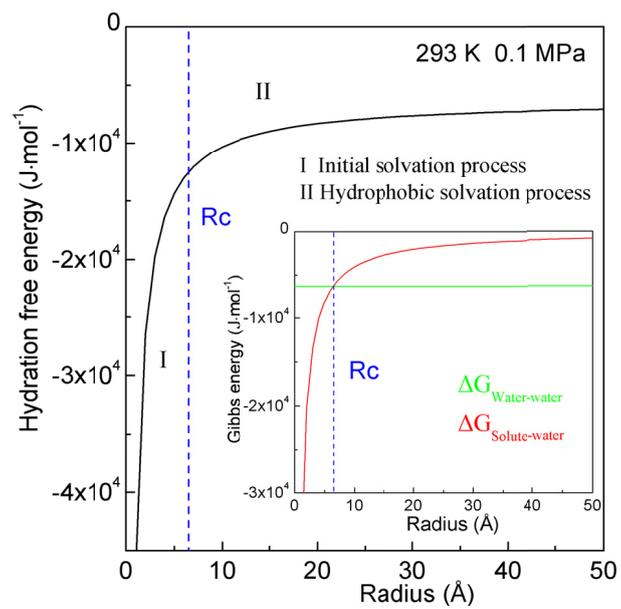

Fig. 1.

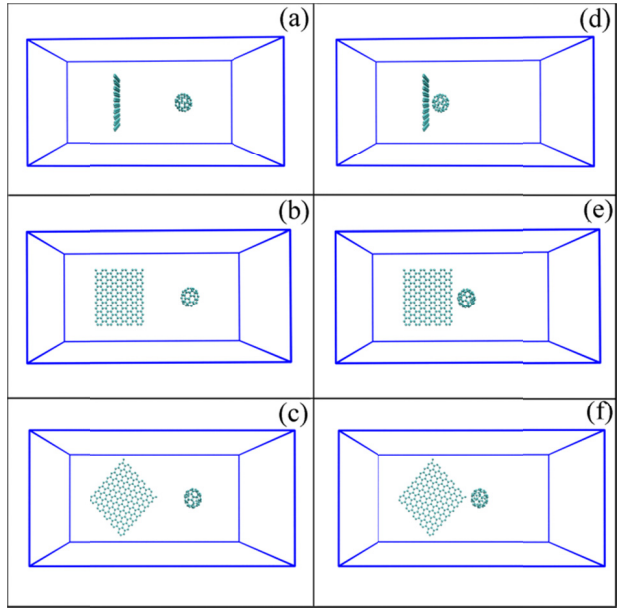

Fig. 2.

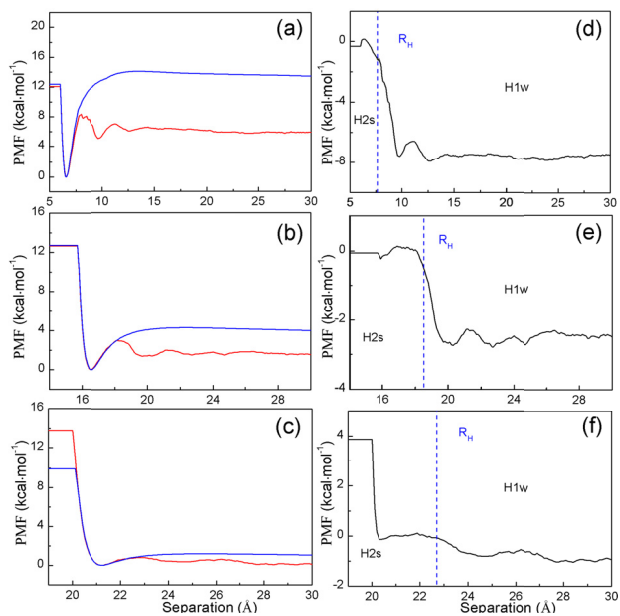

Fig. 3.

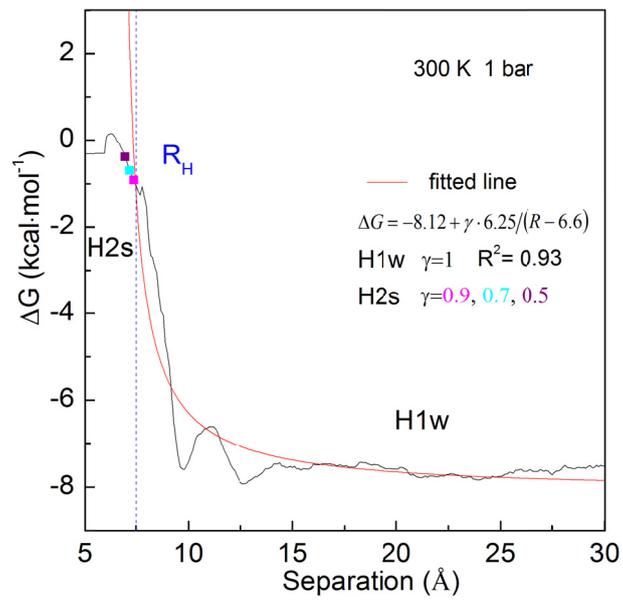

Fig. 4.

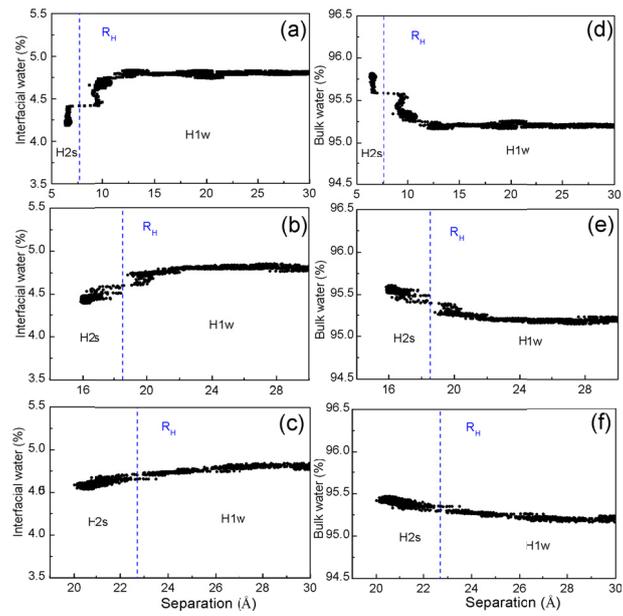

Fig. 5.

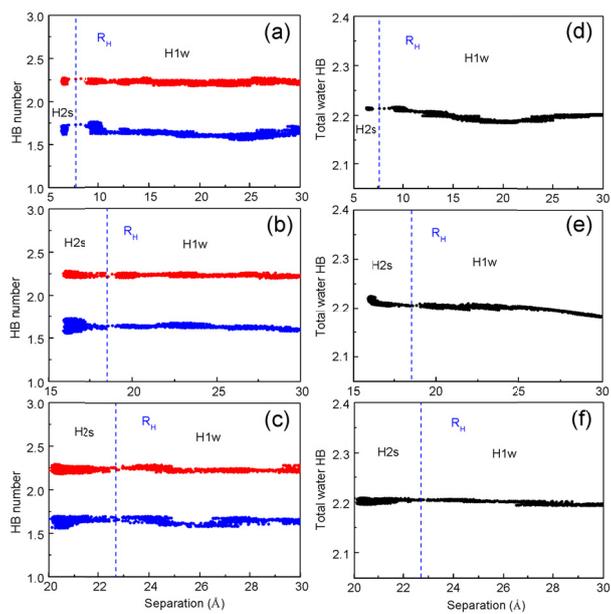

Fig. 6.

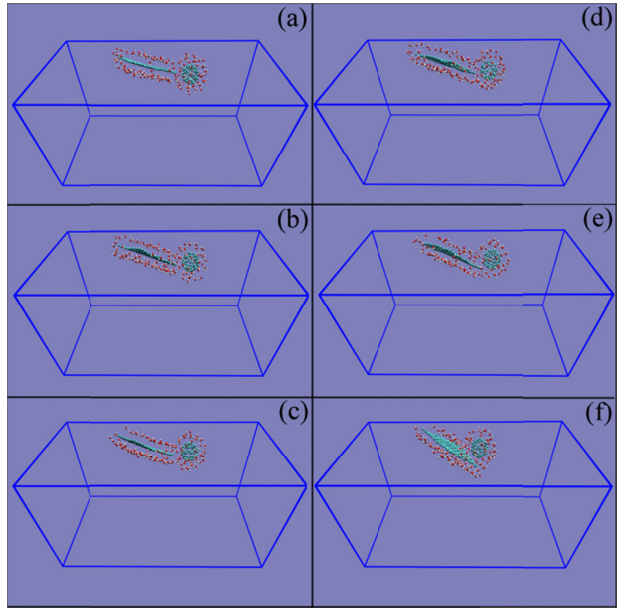

Fig. 7.

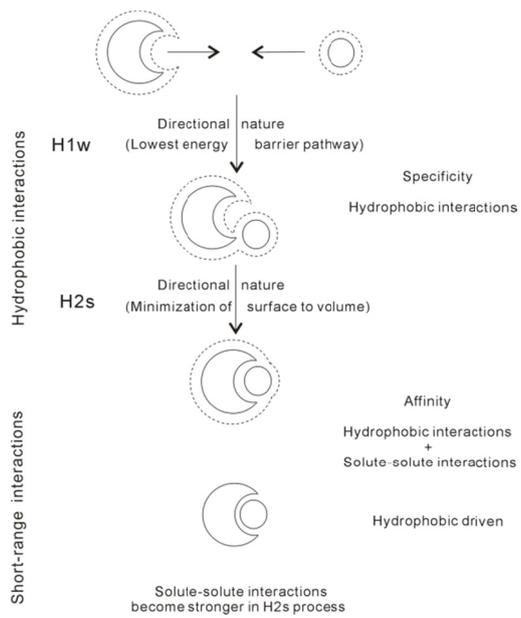

Fig. 8.